\documentclass[
 reprint,
 amsmath,amssymb,
 aps,
 pra,
floatfix,
]{revtex4-2}

\usepackage{graphicx}
\usepackage{dcolumn}
\usepackage{bm}
\usepackage{hyperref}
\usepackage[mathlines]{lineno}

\usepackage{amsmath}
\usepackage{xcolor, ulem}
\begin{document}
\preprint{APS/123-QED}

\title{Study of non-uniformity-induced structured spectrum in a tapered optical fiber through stimulated four-wave mixing}

\author{Minjae Hong}
 \email{minjae.hong@mpl.mpg.de}
\author{Nicolas Y. Joly}%
\affiliation{Department of Physics, Friedrich-Alexander University of Erlangen-Nu\"rnberg, 91058 Erlangen, Germany}
\affiliation{Max Planck Institute for the Science of Light, 91058 Erlangen, Germany}

\date{\today}

\begin{abstract}
Tapered optical fibers offer tunable dispersion and a small mode area, yielding high nonlinearity, which makes them ideal for nonlinear and quantum optics. Because phase-matched four-wave mixing strongly depends on dispersion, the taper diameter becomes a key parameter. For sub-micrometer tapers, identifying the exact diameter is challenging. 
Here, we show that when seeding with a broad spectrum, stimulated four-wave mixing yields a comb-like spectrum that is directly linked to the overall taper profile opening possibility for non-destructive evaluation of sub-micron taper. 
Our model based on joint-spectral amplitude includes the phase accumulated over the fiber length.

\end{abstract}

\maketitle


Photon pair sources have been broadly exploited in various fields of quantum technology, for instance, quantum sensing~\cite{gatto2014beating}, quantum communication~\cite{peng2005experimental}, and photonic quantum computers~\cite{psiquantum2025manufacturable}. Conventionally, photon pairs are generated via spontaneous parametric down-conversion (SPDC) in $\chi^{(2)}$ crystals~\cite{hong2024polarization} and spontaneous four-wave mixing (SFWM) in third-order nonlinear materials, hold in optical fibers~\cite{rarity2005photonic,smith2009photon,lopez2023tunable} and waveguides~\cite{tanzilli2001highly, martin2010polarization}. The benefits of employing optical fibers include their long interaction lengths, which can compensate for the inherently weak nonlinearity of silica. Moreover, for microstructured optical fibers the dispersion properties, essential to phase-matched FWM are engineerable through their geometry~\cite{ReevesNat, smith2009photon, rarity2005photonic}. To adjust the dispersion landscape, aka the phase-matching conditions and hence the frequencies of the generated sidebands, temperature or stress can be used~\cite{ortiz2017spectral}. More flexible however is the use of gases surrounding the solid core of the fiber~\cite{hammer2020broadly}. 
In this context, sub-micrometer tapered fibers (SMTFs) are a particularly efficient platform. SMTFs have a core diameter in the few-hundreds of nanometers, so that the huge evanescent field resides outside the silica core. This unique feature enables control of the dispersion property by adjusting the environmental gas pressure~\cite{hammer2018dispersion}.



The challenge of using SMTFs for generating a pair of photons so-called \textit{signal} and \textit{idler}, is that the required phase matching is directly governed by the taper diameter and the environmental conditions. Any variations of the diameter along the propagation length will lead to deviation from the target frequencies of the pair, which degrades the quality of the source.
Commonly, the source of biphotons is characterized by stimulated emission tomography (SET)~\cite{liscidini2013stimulated}. In SET, the pump beam is combined with a narrowband tunable seed so as to enter the parametric amplification regime~\cite{agrawal2012nonlinear} on the signal side. The location of the peak of the signal indicates the exact phase-matching conditions. Applied to tapered and suspended core fibers, such a technique gives access to the core diameter, which dictates by the dispersion~\cite{hammer2020broadly}. Here we replaced the tunable seed on the idler side with a broadband source.
On the signal side, we observe an unexpected comb-like spectrum, instead of a well-defined single frequency. Additionally, the signal spectrum red-shifts with increasing the environmental gas pressure, while preserving the entire structure. Although random fluctuations of radius on the fiber have been studied~\cite{cui2011spectral}, we show here that this structured spectrum is a direct consequence of the whole shape of the taper. Additionally, we utilize pressure-dependent comb-like fringes to retrieve the overall taper profile in a nondestructive way.
This contrasts with existing nondestructive methods, which do not give a whole taper profile~\cite{wiedemann2010measurement, godet2017brillouin} or offer a limited resolution within a few tens of nanometers~\cite {warken2004fast,azzoune2019optical}.

\bigskip
SFWM is a third-order nonlinear process, where two pump photons annihilate to a pair of a signal photon and an idler photon, respectively at $\omega_s$ and $\omega_i$. The process should obey energy conservation, which imposes that the sidebands are created at equal frequency separation from the pump, with a detuning $\Delta\Omega = \omega_s - \omega_p = \omega_p - \omega_i >0$ that is dictated by phase-matching conditions 
\begin{equation}
    \Delta\beta=2\beta_p - \beta_s - \beta_i -2\gamma P_p=0.
    \label{eq:1}
\end{equation}
There, $\Delta\beta$ is the phase mismatch and $\beta_j$ is the propagation constant of modes ($j=p,s,i$), given by $\beta_j=n_{eff}^{(j)}k_0/c$. $n_{eff}^{(j)}$ is the effective refractive index of the mode at $\omega_j$ which is determined by the fiber geometry, $k_0$ is the vacuum wavenumber and $c$ is the speed of light. $\gamma$ is a nonlinear parameter which takes into account the Kerr nonlinearity~\cite{agrawal2012nonlinear}. For a monochromatic pump with a given power $P_p$ and a homogeneous tapered fiber with a constant waist radius, the wavelengths of the sidebands are well localized and spectrally correlated. 

\begin{figure} [ht]
    \centering
    \includegraphics[width=1.0\linewidth]{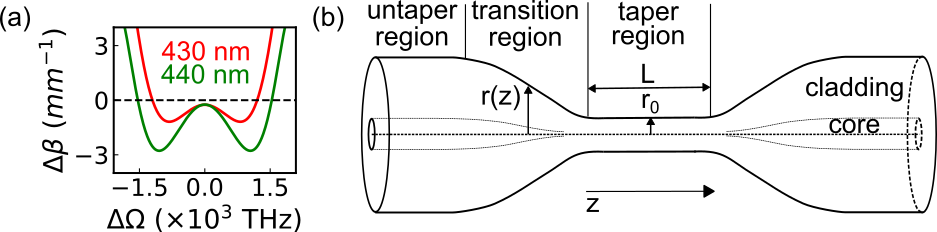}
    \caption{(a) Calculated $\Delta\beta$ using Eq.~\eqref{eq:1} as a function of $\Delta\Omega$ for resp. 430 nm (red) and 440 nm (green) of taper radius. (b) A schematic illustration of a tapered fiber composed of an untaper region, a transition region with varying radius $r(z)$ as a function of fiber position $z$, and a taper region with a taper length $L$ and taper waist $r_0$.
    }
    \label{fig:FWM}
\end{figure}

\begin{figure}[h]
    \centering
    \includegraphics[width=0.9\linewidth]{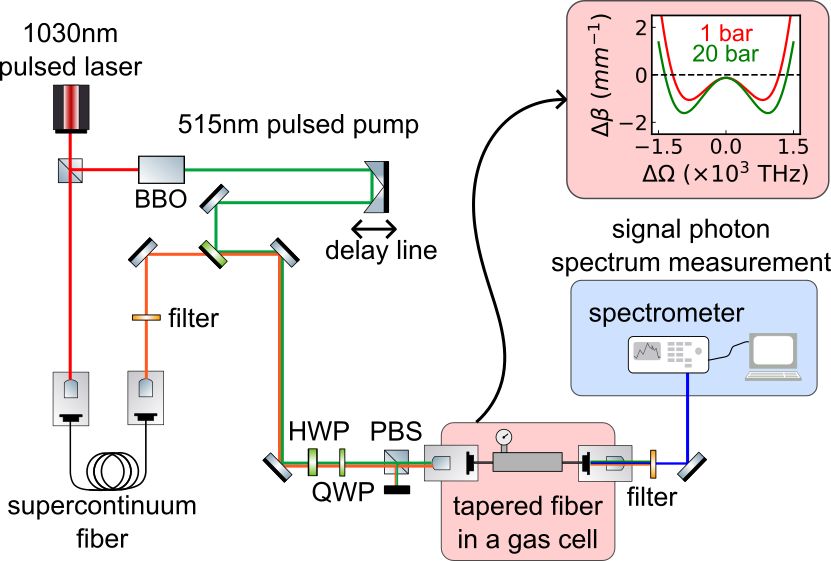}
    \caption{Experimental setup to measure the signal spectrum through stimulated FWM. The 1030 nm laser pumps a PCF to generate a supercontinuum spectrum. The pump is frequency double through a BBO crystal. 
    The pump at 515~nm and the blue side of the SC pulses are overlapped in time using an adjustable delay line. A half-wave plate (HWP), a quarter-wave plate (QWP), and a polarizing beam splitter (PBS) control the pump and seed intensities entering the tapered fiber that is encapsulated in a gas cell. On the upper right side, $\Delta\beta$ as a function of $\Delta\Omega$ is calculated for two different argon pressures.}
    \label{fig:setup}
\end{figure}

\bigskip
Tapered fibers are commonly prepared using flame-brushing technique~\cite{birks1992shape}. The radius in the transition region decreased adiabatically to avoid the coupling of high-order modes and minimize the transmission loss~\cite{love1991tapered, birks1992shape}. Operating at a pump wavelength of 515~nm, we target a tapered waist around 430~nm, leading to a zero-dispersion wavelength ($\lambda_{ZDW} = 520\,$nm) close to the pump wavelength, so as to create signal and idler around 484~nm and 550~nm respectively. As shown in Fig.~\ref{fig:FWM}a, a minute change of the radius waist yields a large shift in the generated frequencies. Such stringent fabrication tolerances can be mitigated with temperature or stress~\cite{ortiz2017spectral} or by encapsulating the taper inside a gas-cell~\cite{hammer2018dispersion} (Fig.~\ref{fig:setup}). 

The experimental setup is depicted in Fig.~\ref{fig:setup}. A Q-switched laser operating at 1030~nm with $\sim2\,$ns of pulse duration and 2~kHz repetition rate pumps a few-meter long photonic crystal fiber (PCF) to generate a broadband supercontinuum (SC)~\cite{wadsworth2004supercontinuum}. On its blue side, the spectrum of the SC extends down to below 530~nm, allowing seeding our FWM process. The pump laser is frequency doubled to 515~nm and coupled into a tapered fiber together with the red-filtered SC (seed). To avoid saturation of our detection scheme, we filter out the 1030~nm pump wavelength using a short-pass filter with a cutoff at $\lambda = 1\mu$m. Any remaining 515~nm pump is filtered out at the end of the tapered fiber. The taper is encapsulated in a gas cell filled with argon gas. The plot on the upper right of Fig.~\ref{fig:setup} presents the phase mismatch for a taper with a fixed radius of 430~nm surrounded by argon at two different pressures. The signal spectrum is measured with a spectrometer (Ocean Optics, Maya 2000 Pro).

\begin{figure}[h]
    \centering
    \includegraphics[width=0.8\linewidth]{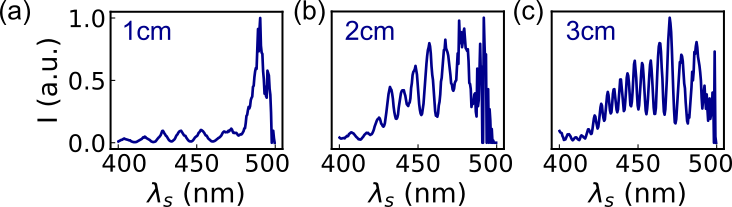}
    \caption{Signal spectrum for (a) 1~cm, (b) 2~cm, and (c) 3~cm  long tapers encapsulated in 5~bar of argon. To remove the influence of the spectral shape of the seed, we normalized the signal spectrum with the intensity of the conjugated wavelength of SC.}
    \label{fig:result}
\end{figure}

\bigskip
We prepared tapered fibers from a commercial step-index solid core fiber (SMF-28)  using a plasma-based tapering machine (3SAE, Tapering Manufacturing Stage). We target a waist radius of 430~nm with different lengths. Figure~\ref{fig:result}~a-c presents the measured signal spectra at 5~bar of environmental argon for three different fiber lengths, 1~cm, 2~cm, and 3~cm. When pumping at 515~nm and seeding on the idler side with a broad spectrum, we expect one single peak on the signal side fulfilling Eq.~\eqref{eq:1}, corresponding to one single fulfilled phase-matching condition, hence one constant diameter over the entire length of the taper. However, each case presents a series of well-defined peaks on the signal side (Fig.~\ref{fig:result}).

To explain the observed comb-like structure of the generated spectrum we should recall that the measured signal spectrum corresponds to the projection of the joint spectral intensity (JSI) along $\omega_s$, where JSI gives the joint probability density of photon pair sources, and is commonly used to characterize the spectral properties of photon pair sources. The JSI can be measured by either conditional coincidence counting by a series of single-shot spectra~\cite{hammer2020broadly} or SET~\cite{liscidini2013stimulated}. 
The JSI is the modulus square of the joint spectral amplitude (JSA). It does not contain any phase information. By contrast, the JSA is a complex-valued probability amplitude function. At the output of a tapered fiber with constant radius, it is given by~\cite{ortiz2017spectral, shmakova2024correlated}:
\begin{equation}
\begin{split}
    F(\omega_s,\omega_i) &= \int \alpha(\omega) \alpha(\omega_s+\omega_i-\omega) \times \\
    &\quad
    \operatorname{sinc}\left[ \frac{L\Delta\beta(\omega,\omega_s,\omega_i)}{2}\right] \times \\
    &\quad
    \exp{\left[{i\frac{L\Delta\beta(\omega,\omega_s,\omega_i)}{2}}\right]} ~d\omega,
    \label{eq:uniform}
\end{split}
\end{equation}
where $L$ is the length of the taper waist and $\alpha(\omega)$ is a spectral shape of the pump, normalized with $\int \left| \alpha(\omega) \right|^2d\omega =1$.




\bigskip
While Eq.~\eqref{eq:uniform} considers a uniform taper waist, the real taper shape does not have a uniform shape, and such inhomogeneity deforms the JSA, hence $\rm{JSI} = \left|\rm{JSA}\right|^2$ of the system~\cite{cui2011spectral}. 

\begin{figure}[h]
    \centering
    \includegraphics[width=1.0\linewidth]{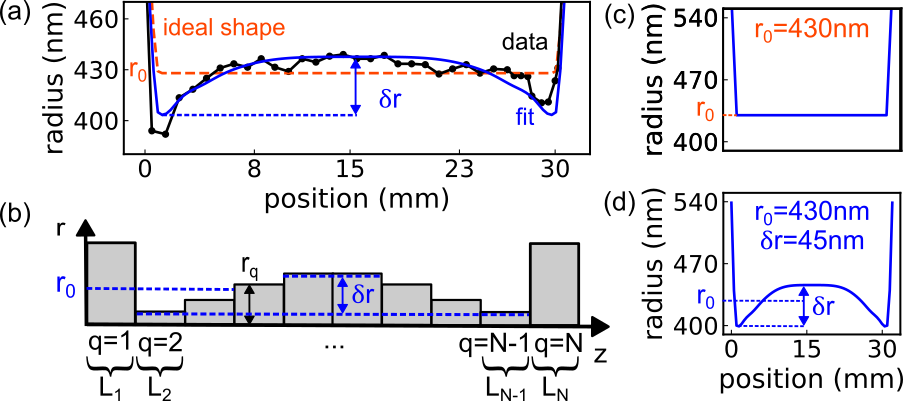}
    \caption{(a) Measured radius at the taper region using scanning electron microscopy is shown in black, and a super-Gaussian fit in blue with a dip size of $\delta r$ with the average radius of $r_0$. The orange curve depicts the ideal shape of the taper region with a uniform radius of $r_0$. (b) Discretization of the taper shape (a) used for the modeling.
   (c) and (d) indicate typical profiles of a taper region for JSA calculation. (c) represents the ideal case, which has a uniform taper radius, and (d) the realistic case, which has a non-uniform radius of super-Gaussian shape. Both (c) and (d) have the same $r_0=430$~nm. In (d), we consider a dip size of $\delta r=45$ nm.}
    \label{fig:shape}
\end{figure}

In the taper manufacturing process, overshooting of the heat source is a common reason for the imperfection, which generates dips at both taper ends~\cite{farwell1997fusedtapered}, yielding a bell-shaped taper profile. In Fig.~\ref{fig:shape}~a, the black curve represents the taper profile of our in-house fabricated SMTF, which is fitted by the super-Gaussian function in the blue curve, while the orange curve indicates the ideal profile with a constant taper radius.

We modeled the shape of the non-uniform taper by radius $r(z)$ varying along the fiber tapered waist,
\begin{align}
    r(z) = r_0-\frac{\delta r}{2} +  \delta r\exp{\left[\left( - \frac{x^2}{2\sigma^2}  \right)^{p}\right]},
    \label{eq:super-Gaussian}
\end{align}
where the dip size $\delta r$ in Fig.~\ref{fig:shape} and $p$ is the power of gaussian function ($p=1$ for a Gaussian shape), $r_0$ the taper radius at full width half maximum and $\sigma$ is the half width half maximum
of a super-Gaussian taper shape. Figure~\ref{fig:shape}b is a discretization of a non-uniform tapered fiber for modeling. Each segment with local length $L_q$ ($q=1,2,3...,N$) has a local radius $r_q$, yielding photon pairs with the frequencies determined by $\Delta\beta_q$. The photon pairs generated in a radius at a certain segment accumulate phases as they propagate along the rest of the taper, eventually leading to interference with the photon pairs generated at the latter segments. Considering this extra phase term, the JSA can be expressed in a form similar to that used in \cite{cui2011spectral}:
\begin{equation}
\begin{split}
    F(\omega_s,\omega_i) &= 
    \sum\limits_{q=1}^N 
    \int \alpha(\omega) \alpha(\omega_s+\omega_i-\omega) \times \\
    &\quad
    \operatorname{sinc}\left[ \frac{L_q\Delta\beta_q(\omega, \omega_s,\omega_i)}{2}\right] \times \\
    &\quad
    \exp{\left[{i\frac{L_q\Delta\beta_q(\omega, \omega_s,\omega_i)}{2}}\right]} \times \\
    &\quad
    \exp{\left[\sum\limits_{n=q+1}^N i \Delta\beta_n(\omega, \omega_s,\omega_i)L_n \right] }  d\omega~.
    \label{eq:nonuniform}
\end{split}
\end{equation}
The last exponential term accounts for the accumulated phase difference between the segment $q$ to the last segment $q=N$.

\begin{figure}[ht]
    \centering
    \includegraphics[width=0.8\linewidth]{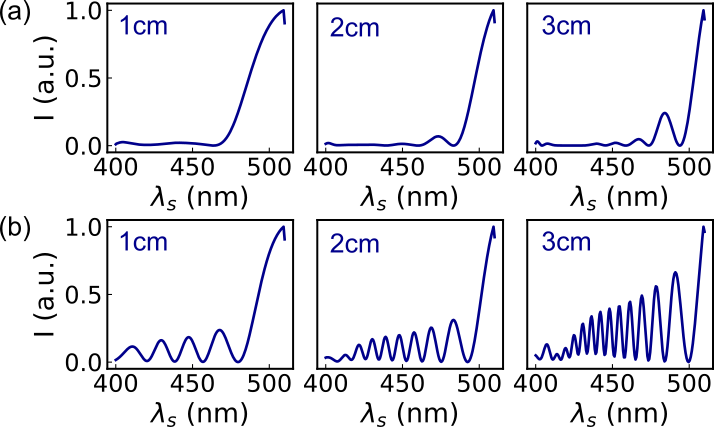}
    \caption{Simulation of signal spectrum of (a) a uniform taper (Fig.~\ref{fig:shape}c) and (b) a non-uniform taper (Fig.~\ref{fig:shape}d) with $L=$ 1 cm, 2 cm, and 3 cm.}
    \label{fig:JSI}
\end{figure}

\bigskip
Equations~\eqref{eq:uniform} and \eqref{eq:nonuniform} are respectively used for calculating the signal spectra of a uniform and a non-uniform tapered fiber. Here, we assumed the pump wavelength $\lambda_p=515$~nm with 1~nm of bandwidth, identical to the experiment. For a non-uniform taper profile, a dip size $\delta r= 45\,$nm is considered based on the taper radius measurement using scanning electron microscopy, and is modeled with Eq.~\eqref{eq:super-Gaussian} with $p=2$.
$\sigma=4$~mm, 8~mm, and 12~mm are considered for 1~cm, 2~cm, and 3~cm taper length, respectively. The complete taper profiles of uniform and non-uniform fibers with $L=3$~cm are presented in Fig.~\ref{fig:shape}c and d.
At a fixed argon pressure of $P=5\,$bar, the signal spectrum $I(\lambda_s)$ of a uniform and non-uniform fiber with $L=$ 1~cm, 2~cm, and 3~cm are calculated and are shown in Fig.~\ref{fig:JSI}a and b, respectively.

As expected for a uniform taper with a constant radius (Fig.~\ref{fig:JSI}~a), the signal spectrum displays a single dominant peak surrounded by smaller high-order peaks of $|\operatorname{sinc}(L\Delta\beta(\omega_s, \omega_i)/2)|^2$: the bandwidth becomes narrower as $L$ increases. On the other hand, comb-like fringes are depicted in the non-uniform tapered fiber (Fig.~\ref{fig:JSI}b). Remarkably, it is clearer to observe the narrower fringes with longer $L$ than in the uniform tapered fiber.
In the non-uniform tapered fiber, photons generated at a given radial position near the front section of the taper accumulate a phase over the propagation and subsequently interfere with photons generated at the same radial position near the rear section of the taper. The fringe shape spectrum is the consequence of varying accumulated phase with local taper radius ($r_q$), because of a super-Gaussian shape of the taper. 

To understand this more in detail, let us assume the two sets of local segments to have equal radii within each set $q=\{1,N\}$ ($r_1=r_N$) and $q=\{2,N-1\}$ ($r_2=r_{N-1}$) (Fig.~\ref{fig:shape}b). Photon pairs $\omega_{s/i,n} (n=1, 2, N-1, N)$ are generated at each segment have the relation: $\omega_{s/i,1} =\omega_{s/i,N}$ and $\omega_{s/i,2} =\omega_{s/i,N-1}$. The photon at $\omega_{s/i, 1}$ will interfere with the photon at $\omega_{s/i, N}=\omega_{s/i, 1}$ after propagating the distance between the segments $q=1$ and $q=N$, which is $z_{1,N}=\sum_{q=1}^{N}L_q$. At the same time, the photon at $\omega_{s/i, 2}$ interfere with the photon at $\omega_{s/i, N-1}=\omega_{s/i, 2}$ after propagating $z_{2,N-1}=\sum_{q=2}^{N-1}L_q$, that is shorter than $z_{1,N}$. Eventually, the optical distance $z_iz_{N-i}$ between two segments $i$ and $N-i$, that generate the same frequencies can either lead to constructive or destructive interferences yielding the comb-line structure of the observed spectrum.  



\begin{figure}[hb!]
    \centering
    \includegraphics[width=0.8\linewidth]{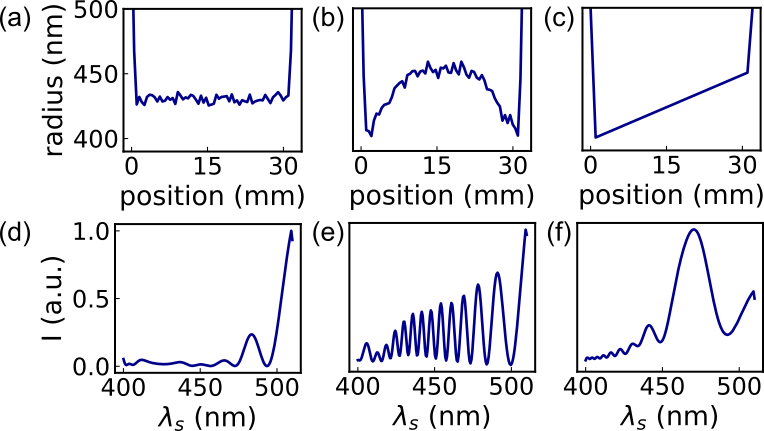}
    \caption{A taper shape with $\pm$3 nm of random fluctuations 
    on (a) the uniform radius and (b) the non-uniform radius, with $L=3$ cm, $r_0=430$ nm, and $\delta r=50$ nm. (c) A linearly increasing taper radius 400 nm to 450 nm. (d)-(f) Corresponding simulated spectral intensity of taper shapes (a)-(c), respectively.}
    \label{fig:JSI_add}
\end{figure}

In order to convince ourselves that the fringes mainly occur by phase accumulation in a super-Gaussian taper profile, which is symmetric along $z$, we considered two different cases: (i) the additional random fluctuation on either a uniform or a non-uniform shape, (ii) a shape in which the same radius does not reappear, for instance, a linearly increasing radius along its length.
Figure~\ref{fig:JSI_add}a and b depict a uniform and a taper with a super-Gaussian profile with an additional $\pm 3\,$nm random fluctuation along its entire length set to $L=3\,$cm. The corresponding calculated spectral densities (resp. Fig.~\ref{fig:JSI_add}d and e) clearly show that the major effect comes from the overall shape of the taper and that fluctuations play a minor role.


In Fig.~\ref{fig:JSI_add}c, we simulate a taper with a linearly increasing radius. 
 In contrast to the case of a super-Gaussian taper profile, such a taper shape produces a comparatively blunt spectral envelope (Fig.~\ref{fig:JSI_add}f). This behavior arises because each taper radii \(r_q\) yield a different signal wavelength, resulting in a broad spectrum but not allowing any interference. 
 This proves that the well-defined comb-like spectrum results from a bell-shape taper profile, in line with the real fiber waist fabricated by flame-brushing technique. 

\bigskip

\begin{figure} [h]
    \centering
    \includegraphics[width=1.0\linewidth]{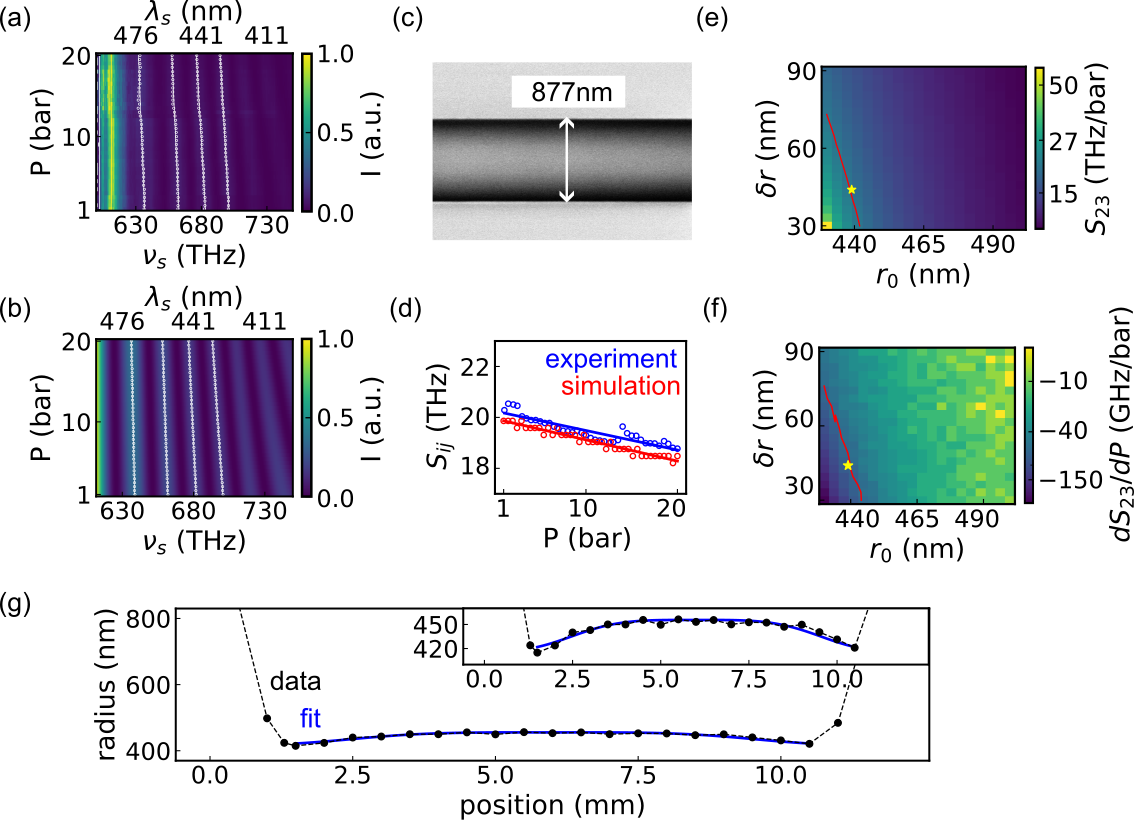}
    \caption{(a) Pressure-tuned signal spectrum measurement of a 1~cm sample. The maxima of fringes are indicated in white circles. (b) Simulated pressure-tuned spectrum of the evaluated taper shape with $r_0=$439~nm, $\delta r=$44~nm, $\sigma=$4~mm, and $p=2$. (c) SEM image of the sample at a certain position. (d) Calculated $S_{23}$ as a function of pressure based on the experimental data (blue) and the evaluated shape of taper (red). Linear fitting is drawn in solid lines.
    (e) Calculated $S_{23}$ map and (f) $dS_{23}/dP$ map with a range of $r_0$ and $\delta r$. Red lines indicate all of the sets ($r_0, \delta r$) satisfying the experimentally found $S_{23}$ and $dS_{23}/dP$ values, and the yellow marker indicates the evaluated $r_0$ and $\delta r$.
    (g) Measured taper profile with SEM. Measurement points are drawn in black, and the super-Gaussian fit is drawn in blue. Vertically enlarged picture of the taper waist is in the upper right.}
    \label{fig:fringe}
\end{figure}

In a second step, we use the structure of the generated spectrum to retrieve the overall profile of the tapered fiber. As shown in Fig.~\ref{fig:fringe}a measured for a 1~cm long taper waist, increasing the environmental argon pressure yields a red-shift of the ensemble of fringes. 
Those fringes are phase-matched, hence strongly linked with dispersion. Since the dispersion both depends on the core diameter and the environmental pressure, monitoring the change of fringe location as a function of pressure allows us to extract the fixed parameter, i.e., the taper profile. Among taper shape parameters in Eq.~\eqref{eq:super-Gaussian}, $r_0$ and $\delta r$ are the major parameters that determine the photon pair wavelength range. 
For this, we use the frequency spacing $S_{ij}$ between two neighboring peaks, $i$ and $j=i+1$, and its pressure dependency $dS_{ij}/dP$.

Here, we focus on $S_{23}$ and $dS_{23}/dP$. From the experiment, we extract $S_{23}^{exp}=20$~THz, $dS_{23}^{exp}/dP=-77$~GHz/bar (Fig.~\ref{fig:fringe}d blue). Numerically, we can compute $S_{23}$ and $dS_{23}/dP$ with a range of $r_0$ and $\delta r$, resulting two 2D-maps. Figure~\ref{fig:fringe}e and f respectively show the $S_{23}$ and $dS_{23}/dP$ maps for a taper with a length $L=9$~mm, $\sigma=4$~mm, and $p=2$. Then, we retrieve the set of ($r_0, \delta r$) that returns the minimum relative error with values $S_{23}^{sim}$ and $dS_{23}^{sim}/dP$, which can be written:
\begin{equation}
    err=\sqrt{\left( \frac{S_{23}^{exp}-S_{23}^{sim}}{S_{23}^{exp}} \right)^2 + \left( \frac{dS_{23}^{exp}/dP-S_{23}^{sim}/dP}{dS_{23}^{exp}/dP} \right)^2}
    \label{eq:error}
\end{equation}
The minimum of the Eq.~\eqref{eq:error} ($err=3.31$\%) gives $r_0=439$~nm and $\delta r=44$~nm. Figure~\ref{fig:fringe}b shows the calculated pressure-dependent structured spectrum. On Fig.~\ref{fig:fringe}d, we plot $dS_{23}/dP$ for the retrieved value. This shows good agreement with the experimental data (\ref{fig:fringe}a and d-blue). We also compared with SEM measurements (Fig.~\ref{fig:fringe}~c) to retrieve the entire profile (Fig.~\ref{fig:fringe}g). Fit of the data with Eq.~\eqref{eq:super-Gaussian} yields $r_0=437$~nm and $\delta r=39$~nm, which is also extremely close to the retrieved values.

\bigskip
In this work, we studied the influence of taper's profile on the generation of phase-matched sidebands through stimulated four-wave mixing.
By seeding with a broadband source on the idler side, we observed well-defined peak structure, whose location depends on the environmental pressure. Considering the phase accumulated along the taper profile (super-Gaussian), we simulated the signal spectrum using JSA. It shows very good agreement with our observation. On the second step, we use the pressure dependence on the location of fringes to evaluate the overall shape of the taper. Although the evaluation requires fixed parameters such as $\sigma$ and $p$ of the super-Gaussian function, this method promises information about the overall shape of SMTF without destroying the fiber.
We believe our approach enables feedback-driven optimization: by analyzing the spectral fingerprint of the FWM process, one can iteratively refine the taper geometry to maximize quantum state generation efficiency.

\begin{acknowledgments}
This work was supported by the Deutsche Forschungsgemeinschaft (DFG, German Research Foundation)–JO 1090/3-2.
\end{acknowledgments}

\nocite{*}

\bibliography{apssamp}

\end{document}